\documentclass[conference,a4paper，10pt]{IEEEtran}
\IEEEoverridecommandlockouts
\usepackage{cite}
\usepackage{bm}
\usepackage[cmex10]{amsmath} 
\usepackage{extarrows}
\usepackage{amssymb}
\usepackage{graphicx}
\usepackage{color}
\usepackage{enumerate}
\usepackage{bookmark} 
\graphicspath{{figure}}
\usepackage{setspace}
\usepackage{subfigure}
\usepackage{algorithm}
\usepackage{algorithmicx}
\usepackage{multirow}
\usepackage{stfloats}
\usepackage{graphbox}
\usepackage{bbm}
\usepackage[utf8]{inputenc} 
\usepackage[T1]{fontenc}
\usepackage{url}
\usepackage{booktabs}
\usepackage{mathrsfs}
\usepackage{graphicx}
\usepackage{mleftright}       
\mleftright                   
\usepackage{framed}  
\usepackage{scalerel}

\newcommand{\diag}{\mr{diag}}

\newcommand{\mr}{\mathrm}

\newcommand{\BE}{\begin{equation}}
\newcommand{\EE}{\end{equation}}
\newcommand{\BS}{\begin{subequations}}
\newcommand{\ES}{\end{subequations}}
\renewcommand{\bf}{\bm}

\newtheorem{theorem}{Theorem}

\newtheorem{assumption}{Assumption}

\newtheorem{definition}{Definition}

\newtheorem{lemma}{Lemma}
\newtheorem{corollary}{Corollary}
\newcommand{\tabincell}[2]{\begin{tabular}{@{}#1@{}}#2\end{tabular}}

\ifCLASSINFOpdf
\graphicspath{{figure/}}

\else

\fi

\begin{document}

\title{Random Modulation: Achieving Asymptotic Replica Optimality over Arbitrary Norm-Bounded and Spectrally Convergent Channel Matrices}
\vspace{-2cm}
\author{%
  \IEEEauthorblockN{Lei Liu\IEEEauthorrefmark{1}\IEEEauthorrefmark{2}, \emph{Senior Member, IEEE}, 
                    Yuhao Chi\IEEEauthorrefmark{2}, \emph{Senior Member, IEEE},
                    and Shunqi Huang\IEEEauthorrefmark{3}, \emph{Student Member, IEEE} 
                    }
  \IEEEauthorblockA{\IEEEauthorrefmark{1}%
                    College of Information Science and Electronic Engineering, Zhejiang University, China}
  \IEEEauthorblockA{\IEEEauthorrefmark{2}%
                    State Key Laboratory of Integrated Services Networks, Xidian University, China}
  \IEEEauthorblockA{\IEEEauthorrefmark{3}%
                    School of Information Science, Japan Institute of Science and Technology, Japan}
E-mail:\{lei\_liu@zju.edu.cn, yhchi@xidian.edu.cn, shunqi.huang@jaist.ac.jp\}

\thanks{This work was supported in part by the National Natural Science Foundation of China (NSFC) under Grants 62394292, 62301485, and 62201424, in part by the Zhejiang Provincial Natural Science Foundation under Grant LZ25F010002, in part by the State Key Laboratory of Integrated Services Networks under Grant ISN25-10, and in part by the Fund HZKY20220550.
}
}

\maketitle
\begin{abstract}
 This paper introduces a random modulation technique that is decoupled from the channel matrix, allowing it to be applied to arbitrary norm-bounded and spectrally convergent channel matrices. The proposed random modulation constructs an equivalent dense and random channel matrix, ensuring that the signals undergo sufficient statistical channel fading. It also guarantees the asymptotic replica maximum \emph{a posteriori} (MAP) bit-error rate (BER) optimality of approximate message passing (AMP)-type detectors for linear systems with arbitrary norm-bounded and spectrally convergent channel matrices when their state evolution has a unique fixed point. Then, a low-complexity cross-domain memory approximate message passing (CD-MAMP) detector is proposed for random modulation, leveraging the sparsity of the time-domain channel and the randomness of the random transform-domain channel. Furthermore, the optimal power allocation schemes are derived to minimize the replica MAP BER and maximize the replica constrained capacity of random-modulated linear systems, assuming the availability of channel state information (CSI) at the transceiver. Numerical results show that the proposed random modulation can achieve BER and block-error rate (BLER) performance gains of up to $2\sim 3$ dB compared to existing OFDM/OTFS/AFDM with 5G-NR LDPC codes, under both average and optimized power allocation. 
\end{abstract}

\section{Introduction} 

As wireless applications rapidly evolve, wireless channels have become more complex, driving the continuous evolution of wireless technology to ensure high-rate high-reliability communications. In 5G communication systems, orthogonal frequency division multiplexing (OFDM) ensures reliable signal transmission in static multipath channels while avoiding inter-symbol interference (ISI). Nevertheless, in emerging wireless applications, such as high-mobility communications (e.g., high-speed railways~\cite{HighSpeed}, low Earth orbit satellites~\cite{LEO}), device mobility causes the wireless channels to be affected by the additional Doppler effect, resulting in doubly selective channels. This compromises the subcarrier orthogonality in OFDM, leading to a substantial decline in performance.

To address this issue, orthogonal time frequency space (OTFS)~\cite{OTFS1} and affine frequency division multiplexing (AFDM) \cite{AFDM} techniques have been proposed in recent years to constructing sparse or nearly sparse equivalent channel matrices to suppress the ISI. Although considerable progress has been made, the development of low-complexity and high-reliability detection algorithms capable of achieving maximum \emph{a posteriori} (MAP) bit-error rate (BER) performance for OTFS and AFDM systems remains an open challenge. State-of-the-art low-complexity and replica-optimal signal recovery algorithms\cite{AMP2009, MaAcess2017, Rangan2019TIT, LeiMAMP, fan2022approximate,zhoufanTIT,Barbier2018ISIT,CAMP}, such as approximate message passing (AMP)~\cite{AMP2009}, orthogonal AMP (OAMP)\cite{MaAcess2017}, vector AMP (VAMP)\cite{Rangan2019TIT}, and memory AMP (MAMP) \cite{LeiMAMP}, are potential solutions. However, their theoretical analyses are generally predicated on assumptions of independent and identically distributed or right-unitarily invariant channel matrices \cite{AMP2009,MaAcess2017,Rangan2019TIT,LeiMAMP,liu2021capacity,LeiOptOAMP,Code_MAMP,zhoufanTIT,Barbier2018ISIT,CAMP,fan2022approximate}. In practical applications, channel distributions often deviate from these assumptions, resulting in performance degradation.

To tackle these challenges, the interleave frequency division multiplexing (IFDM) has been proposed \cite{IFDM}. It utilizes an inverse fast Fourier transform (IFFT) matrix cascaded with a random interleaver, termed the IF transform, to construct a dense and statistically stable equivalent channel matrix, ensuring reliable signal transmission. However, in IFDM, the modulation matrix is limited to the interleaved IFFT. This paper generalizes IFDM to broader random modulation classes, ensuring the equivalent channel matrix lies in the universality class \cite{Rishabh2024}, covering IID, Haar-distributed, and certain permutation-invariant matrices.


In a nutshell, existing modulation and signal detection techniques are heavily dependent on specific assumptions regarding channel matrix structures, such as cyclic Toeplitz matrices of static multipath channels in OFDM\cite{tse2005fundamentals} and single-carrier frequency-domain equalization (SC-FDE) \cite{SC-FDE}, doubly selective channels for OTFS\cite{OTFS1} and AFDM\cite{AFDM}, which substantially restricts their applicability to  complex and dynamic real-world wireless channels. Moreover, \emph{it is shown that in coded systems, OTFS and AFDM do not provide significant performance gains compared to OFDM.}

To address the aforementioned challenges, we introduce a novel random modulation framework that unifies and extends the principles of IFDM \cite{IFDM} and energy-spreading-transform (EST)\cite{EST,EST-EQ}. This framework employs a unitary matrix for modulation, such that the equivalent matrix belongs to the universality class. It can be applied to arbitrary norm-bounded and spectrally convergent channel matrices. The proposed random modulation preserves the performance limits of linear systems, including the replica MAP BER and constrained capacity. To avoid the high complexity of direct signal detection on equivalent dense matrices, a cross-domain MAMP detector has been proposed, which utilizes both the sparsity of the time-domain channel matrices and the property of the equivalent channel matrices. Meanwhile, when the channel state information (CSI) is available at the transceiver, power allocation optimization strategies are proposed for random modulation systems with arbitrary discrete signaling, aiming to achieve the replica MAP BER and maximize the replica constrained capacity.  Numerical results show that the proposed random modulation can achieve BER and block-error rate (BLER) performance gains of up to $2\sim 3$ dB compared to existing OFDM/OTFS/AFDM with 5G-NR LDPC codes, under both average and optimized power allocation. 


Due to the page limit, the proofs of theorems are deferred to the extended version of this paper \cite{RM}. Our code is available at [https://github.com/LeiLiu-s-Lab/Random-Modulation].

\emph{Notation:} $\|\bm{A}\|_{\max} \equiv \max_{i,j} |A_{i,j}|$ denotes the max norm. $f_N \lesssim g_N$ indicates that $f_N \leq C g_N$ for some constants $C > 0$ and large enough $N$.

\section{Random Modulation}

Consider a standard linear system: 
\BE\label{Eqn:linear_sys}
    \bm{y} = \bm{A}\bm{x} + \bm{n},
\EE
where $\bm{y} \in \mathbb{C}^{M}$ is a vector of observations, $\bf{A} \in \mathbb{C}^{M \times N}$ is a given channel matrix,  $\bf{x} \in \mathbb{C}^{N}$ is a vector to be estimated, and $\bm{n} \sim \mathcal{CN}(\mathbf{0},\sigma^2\bm{I}_M)$ is a Gaussian noise vector. Let ${\rm snr} = \sigma^{-2}$ represent the transmit signal noise ratio (SNR).

\begin{definition}
    We say that a matrix $\bm{A}$ is \emph{spectrally convergent} if the empirical spectral distribution of $\bm{A}^{\mr{H}}\bm{A}$ converges to a compactly supported probability distribution on $[0, \infty)$.
\end{definition}
\begin{assumption}\label{ASS:Model}
The average power of $\bf{x}$ is normalized, i.e., {$\lim_{N\to\infty} \frac{1}{N}\|\bf{x}\|^2\overset{\rm a.s.}{=}1$}. Consider a large-scale linear system that $M,N\to\infty$ with a fixed $\delta=M/N$. The matrix $\bf{A}$ is spectrally convergent, and its spectral norm is bounded by a constant, i.e., $\|\bf{A}\|_2 \lesssim 1$.
\end{assumption}

The vector $\bm{x}$ is obtained by modulating an i.i.d. signal vector $\bm{s} \in \mathbb{C}^N$ through a transform matrix $\bm{\Xi} \in \mathbb{C}^{N \times N}$, i.e., $\bm{x} = \bm{\Xi} \bm{s}$. Hence, the system in \eqref{Eqn:linear_sys} can be written as:
\BE\label{Eqn:RUP_sys}
    \bm{y} = \bm{A}\underbrace{\bm{\Xi s}}_{\bm{x}} + \bm{n}, \quad
    {\rm s.t.}\ s_i\sim P_{s}, \; \forall i,  
\EE
where $P_s$ is assumed to be symmetric about the origin (e.g., PAM, PSK, QAM). Subsequently, we formally introduce the random modulation.
\begin{definition}[Random Modulation]\label{Def:RM} $\bf{\Xi} \bf{s}$ is said to be a random modulation of the signal $\bf{s}$ if:
\begin{enumerate}
    \item $\bm{\Xi}$ is a random unitary matrix  satisfying $\bm{\Xi}^{\rm H}\bm{\Xi} = \bm{I}_N$  and independent of $\{\bf{A}, \bf{s},\bf{n}\}$. 
    \item The equivalent channel matrix $\bm{J} \equiv \bm{A\Xi}$ belongs to the universality class $\mathscr{U}$ \cite{Rishabh2024}. That is, 
    \begin{itemize}
        \item $\bm{J}$ is spectrally convergent, and  $\|\bm{J}\|_2 \lesssim 1$.
    \item  For any fixed $k \in \mathbb{N}^*$, $\epsilon > 0$, 
    \BE \label{eq:univ-class-moment-convergence}
        \Big\| (\bm{J}^{\rm H}\bm{J})^k - \frac{\mr{ tr}[(\bm{J}^{\rm H}\bm{J})^k]}{N} \bf{I}_N \Big\|_{\max} \lesssim N^{-1/2+\epsilon}.
    \EE  
    \end{itemize}
\end{enumerate}

\emph{Remark:} The definition of $\mathscr{U}$ in \cite{Rishabh2024} includes a post-multiplication on $\bm{J}$ by a random sign matrix $\bm{B} = \mathrm{diag}(\bm{b})$, where $\bm{b} \overset{\text{i.i.d.}}{\sim} \mathrm{Unif}(\{\pm1\})$. We omit this sign matrix since $P_s$ is assumed symmetric about the origin. In addition, though \cite{Rishabh2024} concentrates on real matrices, we do not distinguish between real and complex cases in this work.
\end{definition}

\begin{figure}[t!]
    \centering
    \includegraphics[width = 0.35\textwidth]{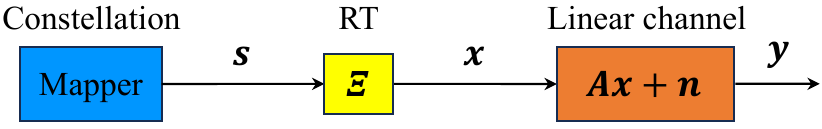}\vspace{-0.3cm}
    \caption{The linear system with random modulation.}\vspace{-0.2cm}
    \label{fig:RUP_model}
\end{figure}

Fig. \ref{fig:RUP_model} illustrates the linear system with random modulation. We denote $\bm{\Xi}$ and $\bm{\Xi}^{\rm H}$ as the random transform (RT) and inverse RT (IRT) matrices. In the following theorem, we summarize commonly used classes of RT matrices. 

\begin{theorem}\label{The:RT}
    Suppose that $\bf{A}$ is spectrally convergent, and its spectral norm is bounded by a constant, i.e., $\|\bm{A}\|_2 \lesssim 1$. Then, $\bm{A\Xi}\in \mathscr{U}$, where the random transform matrix $\bf{\Xi}$ can be
    \begin{enumerate} 
    \item \emph{IID Matrices} \cite[Lemma 4]{Rishabh2024}: $\bf{\Xi}_{\rm IID}$ with IID entries satisfying ${\rm E}\{\Xi_{i,j}\}=0$ and ${\rm E}\{\Xi_{i,j}^2\}= {1}/{N}$. Note that $\bf{\Xi}_{\rm IID}$ is an approximate unitary matrix as $N \to \infty$.
    \item \emph{Haar Distributed Matrices} \cite{Rangan2019TIT, takeuchi2020rigorous, Rishabh2024}: $\bf{\bf{\Xi}}_{\rm Haar}\sim {\rm Unif}(\mathbb{U}(N))$, where $\mathbb{U}(N)$ denotes the space of all $N\times N$ unitary matrices. 
    \item \emph{Permutation-Invariant Matrices}: $\bf{\Xi}_{\mr{PI}} = \bf{\Pi}\bf{U}$, where $\bf{\Pi}$ is a uniformly random permutation matrix, and $\bf{U}$ is a deterministic unitary matrix satisfying $\|\bf{U}\|_{\max}\lesssim N^{-1/2+\epsilon}$ for any $\epsilon>0$. It suffices to assume that $\big|\sum_{i \neq j} [(\bm{A}^{\rm H} \bm{A})^k]_{i,j}\big| \lesssim N^{1/2 + \epsilon}$ for any fixed $k \in \mathbb{N}^*$ and $\epsilon > 0$, though we conjecture that this condition may be unnecessary. In particular, $\bf{U}$ can be selected as a structured fast transform, such as the discrete Fourier transform (DFT), the Hadamard-Walsh transform (HWT), and the interleaved block-sparse transform (IBST) \cite{IBST}. 
\end{enumerate}
\end{theorem}

\section{Cross-Domain Bayes-Optimal MAMP Detector}
\subsection{CD-MAMP Detector}
Rewrite the random modulation system in \eqref{Eqn:RUP_sys} as
\BS\label{Eqn:RUP_dect}
\begin{align}
    \text{Linear constraint}\; \Gamma: \; & \bf{y}=\bf{A}\bf{x}+\bf{n}, \\ 
    \text{Random Transform}\; T:\; &\bf{x}=\bf{\Xi}\bf{s}, \\
    \text{Nonlinear constraint}\; \Phi: \; & \bf{s} \sim P_S(\bf{s}).
\end{align}
\ES

\begin{figure}[t!]
    \centering
    \includegraphics[width = 0.35\textwidth]{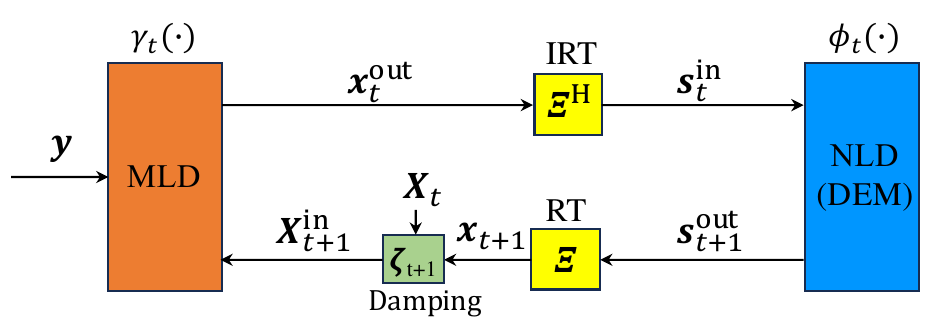}\vspace{-0.3cm}
    \caption{The CD-MAMP detector for the random modulation linear system.}\vspace{-0.2cm}
    \label{fig:BO-MAMP}
\end{figure}

Note that for arbitrary discrete input signaling $\bf{s}$ and general $\bf{A}$, the optimal solution is in general NP-hard \cite{verdu1984optimum}. To address this difficulty, a CD-MAMP is proposed that utilizes both the sparsity of $\bf{A}$ and the fact that  $\bf{A\Xi}$ belongs to the universality class $\mathscr{U}$, as shown Fig.~\ref{fig:BO-MAMP}. The detailed process is given as: Starting with $t=1$, $\bf{X}_1^{\rm{in}}=[\bf{x}^{\rm in}_{1}, ..., \bf{x}^{\rm in}_{t}]=\bf{0}$, and $\bf{s}_1^{\rm out}=\bf{0}$,
\BS\label{Eqn:MAMP}
\begin{align}
\!\!\!\!{\text {\small MLD}\!:}&\; \bf{x}^{\rm out}_t  = \gamma_t(\bf{X}_t^{\rm{in}}) = \tfrac{1}{{\epsilon}^\gamma_t} \left( \bf{A}^{\rm H}{\hat{\bf{r}}}_{t} - {\bf{\mathcal{P}}}_{t}\bf{X}_t^{\rm{in}} \right),\label{Eqn:MLE}\\%
\!\!\!\!{\text {\small IRT}\!:}& \; \bf{s}^{\rm in}_t =\bf{\Xi}^{\rm H}\bf{x}^{\rm out}_t, \label{Eqn:IRUT} \\
\!\!\!\!\!\!\!\!{\text {\small NLD}\!:}& \; \bf{s}^{\rm out}_{t+1}=\phi_t(\bf{s}^{\rm in}_t),\label{Eqn:NLE}\\ 
\!\!{\text {\small RT}\!:}& \; \bf{x}_{t+1} =\bf{\Xi}\bf{s}^{\rm out}_{t+1}, \label{Eqn:RUT}\\
\!\!\!\!\!{\text {\small Damping}\!:}&\; \bf{x}^{\rm in}_{t+1}  \!=\! \bar{\phi}_t(\bf{x}_{t+1})\!=\! [\bf{x}_1^{\rm{in}},\cdots\!,\bf{x}^{\rm{in}}_{t}, \bf{x}_{t+1}]\cdot \bf{\zeta}_{t+1},   \label{Eqn:damp}
\end{align}
\ES
where $\hat{\bf{r}}_t = \bf{B}_t \hat{\bf{r}}_{t-1} + \xi_t(\bf{y}-\bf{Ax}_t^{\rm in})$, $\bf{B}_t = \theta_t({\lambda^\dagger\bf{I}-\bf{AA}^{\rm H}})$ with ${\lambda}^\dag=[\lambda_{\max}+\lambda_{\min}]/2$, $\lambda_{\min}$ and $\lambda_{\max}$ denote the minimal and maximal eigenvalues of $\bf{A}\bf{A}^{\rm H}$, respectively. Meanwhile, $\{\bf{\zeta}_{t+1}, \theta_t,\xi_t, \epsilon_t^{\gamma}\}$ are optimized to ensure replica MAP optimality of CD-MAMP. (See details in \cite{LeiMAMP})

To demonstrate the advantages of CD-MAMP in complexity, we present a comparison with existing state-of-the-art detectors, in which the channel matrix $\bf{A}$ is assumed to be sparse and the number of non-zero elements per row in $\bf{A}$ is $\mathcal{K}$ ($\mathcal{K}\ll {\rm{min}}\{M,N\}$),  e.g., time-varying multipath channels \cite{tse2005fundamentals}, etc. 
Table~\ref{Tab:complexity} presents the comparisons in time and space complexity of CD-MAMP, CD-OAMP~\cite{OTFS-OAMP,TurboCS,EST-EQ}, symbol domain (SD) MAMP~\cite{MAMPOTFSconf}, and SD Gaussian message passing (GMP)~\cite{OTFS_GMP}. Hence,  CD-OAMP, SD-GMP, and SD-MAMP have higher complexity than CD-MAMP for the maximum iteration number $\mathcal{T}\ll N$.
\vspace{-0.4cm}
\begin{table}[h!] \tiny
\renewcommand{\arraystretch}{1.1} 
\caption{Complexity Comparisons of Advanced Detectors}
\label{Tab:complexity} 
\centering \scriptsize \setlength{\tabcolsep}{0.8mm}{
\begin{tabular}{c||c|c}
\hline
Algorithms & Time complexity & Space complexity   \\
\hline 
\hline 
\tabincell{c}{SD-GMP\!\!\vspace{-0.1cm}\\ \cite{OTFS_GMP}} &  $\mathcal{O}(N^2\mathcal{T})$ &$\mathcal{O}(MN)$  \\ 
\hline 
\tabincell{c}{CD-OAMP\!\!\vspace{-0.1cm}\\ \cite{OTFS-OAMP,TurboCS,EST-EQ}}  &  $\mathcal{O}((M^2N\!\!+\!\!M^3)\mathcal{T}\!\!+\!\!2N\mathcal{T}{\mr{log}}N\!)\!$ &  $\mathcal{O}(MN)$ \\
\hline 
\tabincell{c}{SD-MAMP\!\!\vspace{-0.1cm}\\ \cite{MAMPOTFSconf}}  &  $\mathcal{O}(N^2\mathcal{T})$ &    $\mathcal{O}(MN\!+\!M\mathcal{T}\!+\!\mathcal{T}^2)$\\
\hline 
\tabincell{c}{CD-MAMP\!\!\vspace{-0.1cm}\\  (proposed)}   & $\mathcal{O}(\mathcal{K}N\mathcal{T}+2N\mathcal{T}{\mr{log}}N)$ &  $\mathcal{O}(\mathcal{K}M\!+\!M\mathcal{T}\!+\!\mathcal{T}^2)$ \\
\hline 
\end{tabular}}
\end{table}

\vspace{-0.4cm}
\subsection{State Evolution}
According to the asymptotic IID Gaussianity presented in \cite[Lemma 4]{LeiMAMP}, the output covariances of $\gamma_t(\cdot)$ and $\bar{\phi}_t(\cdot)$ in \eqref{Eqn:MAMP} can be evaluated by: As $N\to \infty$,
\BS\label{Eqn:IIDG}
\begin{align}
v_{t,t'}^{\gamma}\! &\!\overset{\rm a.s.}{=}\! \tfrac{1}{N}{\rm E}\Big\{\big[\gamma_t\big(\bf{X}\! +\!\bf{G}_{t})\!-\!\bf{x}\big]^{\rm H}\big[\gamma_{t'}\big(\bf{X}\!+\!\bf{G}_{t'}\big)\!-\!\bf{x}\big]\Big\},\\
 v_{t,t'}^{\bar{\phi}}\!&\!\overset{\rm a.s.}{=}\! 
\tfrac{1}{N}{\rm E}\Big\{\big[\bar{\phi}_t\big(\bf{X}\!+\!\bf{Z}_{t} )\!-\bf{x}\big]^{\rm H} \big[\bar{\phi}_{t'}\big(\bf{X}\!+\!\bf{Z}_{t'}\big)\!-\!\bf{x}\big]\Big\},
\end{align}
\ES
where $\bf{X}=\bf{x}\cdot \bf{1}^{\rm{T}}$ with an all-ones vector $\bf{1}$ of proper size, $\bf{G}_t=[\bf{g}_1,\cdots,\bf{g}_t]$ and $\bf{Z}_t=[\bf{z}_1,\cdots,\bf{z}_t]$ are column-wise IID Gaussian and row-wise joint-Gaussian matrices and independent of $\bf{x}$. Moreover, $\bf{g}_t\sim \mathcal{CN}(0, v_{t,t}^{\gamma}\bf{I})$ with ${\rm E}\{\bf{g}_t(\bf{g}_{t'})^{\rm H}\}\!=\!v_{t,t'}^{\gamma}\bf{I}$ and $\bf{z}_t\!\!\sim\!\mathcal{CN}(\bf{0},v_{t,t}^{\bar{\phi}}\bf{I})$ with ${\rm E}\{\bf{z}_t(\bf{z}_{t'})^{\rm H}\}\!=\!v_{t,t'}^{\bar{\phi}}\bf{I}$. Let $\bf{v}_t^{\gamma}=[v_{t,1}^{\gamma}, ..., v_{t,t}^{\gamma}]^{\rm{T}}$, $\bf{v}_t^{\bar{\phi}}=[v_{t,1}^{\bar{\phi}}, ..., v_{t,t}^{\bar{\phi}}]^{\rm{T}}$, $\bf{V}_t^{\gamma}\equiv[v_{i,j}^{\gamma}]_{t\times t}$ and $\bf{V}_t^{\bar{\phi}}\equiv[v_{i,j}^{\bar{\phi}}]_{t\times t}$. Similar as \cite[Proposition 4]{LeiMAMP}, the state evolution (SE) of CD-MAMP in \eqref{Eqn:IIDG} is given as follows: 
\BE\label{Eqn:MAMP_SE0}
    \bf{v}_t^{\gamma}=\gamma_{\text{SE}}(\bf{V}_t^{\bar{\phi}}), \quad
    \bf{v}_{t+1}^{\bar{\phi}}=\phi_{\text{SE}}(\bf{V}_t^{\gamma}), 
\EE
where $\gamma_{\text{SE}}(\cdot)$ and $\phi_{\text{SE}}(\cdot)$ denote the mean squared error (MSE) functions of $\gamma_t(\cdot)$ and $\bar{\phi}_t(\cdot)$ in \eqref{Eqn:MAMP}, respectively. The specific expressions of \eqref{Eqn:MAMP_SE0} can be found in \cite[Equations (46) and (47)]{LeiMAMP}.

Note that the high-dimensional covariance matrices $\bf{V}_t^{\gamma}$ and $\bf{V}_t^{\phi}$ in \eqref{Eqn:MAMP_SE0} complicate direct application in performance analysis and optimization for random modulation systems. To address this challenge, following $\bf{A\Xi} \in \mathscr{U}$ and \cite[Lemma 3]{Code_MAMP}, we simplify the complex covariance-based SE analysis of CD-MAMP for fixed-point analysis and achievable rate analysis below by using the scalar variance-based SE of CD-OAMP similar as \cite[Equation(28)]{LeiOptOAMP}. (See \eqref{Eqn:iterSEb}).

More importantly, the replica MAP-BER optimality of the CD-MAMP and CD-OAMP algorithms is presented in Corollary~\ref{cor:MAP_BER} based on the following assumption.
\begin{assumption}[Unique SE Fixed Point]\label{Asp:SEfixed} 
Assume that there is a unique fixed point in the SE of CD-MAMP and CD-OAMP.
\end{assumption}

\begin{corollary}[Replica MAP-BER Optimality \cite{Rishabh2024}]\label{cor:MAP_BER}
Suppose that the Assumptions \ref{ASS:Model} and \ref{Asp:SEfixed} hold. In the linear systems with random modulation satisfying $\bf{A}\bf{\Xi} \in \mathscr{U}$, both the CD-OAMP and low-complexity CD-MAMP detectors converge to the replica MAP BER.
\end{corollary}


\section{Power Allocation}

In this section, we study the RT-domain power allocation in linear systems with random modulation, given by
\BE
    \bm{y}=\bm{AP\Xi}\bm{s} + \bm{n}, \label{Eqn:GPA}
\EE
where $\bm{P} \in \mathbb{C}^{N \times N}$ is a power allocation matrix subject to ${\rm tr}[\bm{P}\bm{P}^{\rm H}]=P_{\rm sum}$ (total transmit power).
\begin{theorem}\label{The:Opt_P}
    Let $\bf{A}=\bm{U}_A \bm{\Sigma}_{A} \bm{V}^{\rm H}_A$ and $\bf{P}=\bm{U}_P \bm{\Sigma}_{P} \bm{V}^{\rm H}_P$ be the singular value decomposition of $\bm{A}$ and $\bm{P}$, respectively. Then, regardless of whether the objective is to minimize the MAP BER or to maximize the constrained capacity of the system in \eqref{Eqn:GPA}, it is optimal to set
    \BE\label{Eqn:P_VS}
        \bm{U}_P = \bm{V}_A,
    \EE
    with $\bm{V}_P$ being arbitrary. Without loss of generality, we can set $\bm{V}_P = \bm{I}_N$ such that
    \BE
        \bm{P} = \bm{V}_A\bm{\Sigma}_P.
    \EE
\end{theorem}

Following Theorem \ref{The:Opt_P}, we reformulate \eqref{Eqn:GPA} as 
\begin{align} \label{Eqn:pa1}
    \bm{y} &= \bm{A}\bm{V}_A\bm{\Sigma}_P\bm{\Xi}\bm{s} + \bm{n} \nonumber \\
    &=\bm{U}_A \bm{\Sigma}_{A} \bm{\Sigma}_P \bm{\Xi}\bm{s} + \bm{n},
\end{align}
where $\bm{\Sigma}_{P} = \diag\{\sqrt{p_1}, \cdots\!, \sqrt{p_N}\}$ is the power allocation matrix subject to $\sum_{i=1}^N p_i = P_{\rm sum}$. Next, we rewrite \eqref{Eqn:pa1} as
\BE \label{Eqn:pa2}
    \tilde{\bm y} = \tilde{\bm{\Sigma}}_{A}  \bm{\Sigma}_P \bm {\Xi} {\bm s}  + \tilde{\bm n},
\EE
where $\tilde{\bm y}=\bf{U}_A^{\rm{H}}\bf{y}$, $\tilde{\bm n}=\bf{U}_A^{\rm{H}}\bf{n}$, $\tilde{\bm n} \!\sim\!\mathcal{CN}(\mathbf{0},\sigma^2\bm{I}_N)$, and $\tilde{\bm{\Sigma}}_{A}  ={\rm diag}\{\sigma_1,\dots, \sigma_N\}$ with $\sigma_i = 0$ for $\min\{M,N\}<i\leq N$. As a result, power allocation reduces to optimizing $\bm{p} \equiv [p_1, \cdots\!, p_N]$ to minimize MAP BER or maximize constrained capacity of the system in \eqref{Eqn:pa2}.

\emph{Remark:} In \cite{EST-EQ}, it was proven that \eqref{Eqn:P_VS} is optimal for minimizing $P_{\rm sum}$ when the MAP BER is given, which differs from Theorem \ref{The:Opt_P}. Our proof builds on an intermediate result from their analysis, but the overall approach remains distinct.

According to Theorem~\ref{The:RT}, the equivalent channel matrix $\tilde{\bm{\Sigma}}_{A}\bm{\Sigma}_P \bm{\Xi}$ belongs to the universality class $\mathscr{U}$. This leads to the following corollary.

\begin{corollary}[Replica MAP BER Optimality]\label{The:MAMP_opt_prob}
Suppose that Assumptions \ref{ASS:Model} and \ref{Asp:SEfixed} hold. Then, given the optimal $\bm p^*$, the replica MAP BER in \eqref{Eqn:pa2} can be achieved by both the CD-OAMP detector and the proposed low-complexity CD-MAMP detector in \eqref{Eqn:MAMP}, with the input matrix replaced by $\tilde{\bm{\Sigma}}_{A}\bm{\Sigma}_P$. 
\end{corollary}

\subsection{Power Allocation to Minimize MAP BER}\label{Sec:PA_BER}
Power allocation to minimize MAP BER is tailored for the receivers employing a detector cascaded by a decoder. In this context, effective single-input single-output (SISO)-AWGN codes are sufficient, and maximizing the achievable rate is equivalent to minimizing the MAP BER in detection.

\subsubsection{Problem Formulation} The MAP BER of the linear system in \eqref{Eqn:pa2} can be evaluated by the following lemma.

\begin{lemma}[Replica MMSE and MAP BER]\label{Lem:MMSE_PA}
    The replica MMSE $v^*$ of the linear system in \eqref{Eqn:pa2} is given by  
\BE\label{Eqn:replicaMMSE_PA}
     {\rm mmse}^{-1}(v^*) = {\sigma^{-2}} \cdot \mathcal{R}_{\bf{R}}\left( -\sigma^{-2}v^* \right), 
\EE
where $\mathcal{R}_{\bf{R}}(\cdot)$ is the R-transform with $\bf{R}=\bm{\Sigma}_P^{\rm H}\tilde{\bm{\Sigma}}_{A}^{\rm H}\tilde{\bm{\Sigma}}_{A} \bm{\Sigma}_P$ \cite{Tulino2004}, $\rho={\rm mmse}^{-1}(v)$ is the inverse of $v={\rm mmse}\{\bf{x}|\sqrt{\rho}\bf{x}+\bf{z}, \bf{x}\sim P_{\bf{x}}\}$ with noise $\bf{z}\sim \mathcal{CN}(\bf{0},\bf{I})$ independent of $\bf{x}$ and an effective signal-to-noise ratio $\rho$, ${\rm mmse}\{\bf{x}|\sqrt{\rho}\bf{x}+\bf{z}, \bf{x}\sim P_{\bf{x}}\}\equiv\frac{1}{N}{\mr{E}}\{||\hat{\bf{x}}_{\text{post}}-\bf{x}||^2\}$
 with the {a-posteriori} mean $\hat{\bf{x}}_{\text{post}}={\mr{E}}\{\bf{x}|\sqrt{\rho}\bf{x}+\bf{z}, \bf{x}\sim P_{\bf{x}}\}$.
The replica MAP BER of the linear system in \eqref{Eqn:pa2} is $
    {\rm BER}^*(\bm p) =Q_{\mathcal{S}}\big(\rho^*(\bm p)\big),
$
where $Q_{\mathcal{S}}(\rho)$ denotes the MAP demodulation BER function.
\end{lemma} 

Since $Q_{\mathcal{S}}(\cdot)$ is a monotonically decreasing function, to minimize the MAP BER of the system in \eqref{Eqn:pa2}, the power allocation is formulated as following:
\BS\label{Eqn:MAP_BER_11}\begin{align}
     {\mathcal P}_{1.1}:  \;\;  &\mathop{\rm{max}}_{\bm{p}}  \;\;  \rho^*(\bm p), \\
     & {\rm s.t.}\quad  \textstyle\sum_{i=1}^N p_i = P_{\rm sum}, \\ 
     & \;\;\quad\quad p_i\geq0,\; i=1,\cdots, N.
\end{align} \ES

\subsubsection{Problem Transformation} 
In general, obtaining $\rho^*$ by directly solving \eqref{Eqn:MAP_BER_11} is challenging. Fortunately, as shown in Corollary~\ref{The:MAMP_opt_prob}, CD-OAMP can achieve the replica MAP-BER optimality, and its convergence has been established in~\cite{LeiMAMP, LMOAMP}. Therefore, the SE of CD-OAMP is employed, i.e., 
\BS\label{Eqn:iterSEb}\begin{align} 
    \rho_t^{\gamma} &= {\gamma}_{\mr{SE}}(v_t^{\phi},\bm{p})=\big[\hat{\gamma}_{\rm SE}(v_t^{\phi},\bm{p})\big]^{-1} - [v_t^{\phi}]^{-1}, \label{Eqn:iterSEa}\\
    v_{t+1}^{\phi}&={\phi}_{\mr{SE}}(\rho_t^{\gamma})=\big(\big[{\rm mmse}(\rho_t^{\gamma})\big]^{-1} - \rho_t^{\gamma}\big)^{-1},\label{Eqn:phi_se}
\end{align}
\ES
where
\BE
    \!\!\!\!\hat{\gamma}_{\rm SE}(v_t^{\phi},\bm{p})=\tfrac{1}{N}\mr{tr}\big\{\big([v_t^{\phi}]^{-1}\bf{I}\!+\!\sigma^{-2}\bm{\Sigma}_P^{\mr{H}}\tilde{\bm{\Sigma}}_{A}^{\rm H}\tilde{\bm{\Sigma}}_{A}\bm{\Sigma}_P\big)^{-1}\big\}.\label{Eqn:iterSEc} 
\EE
Hence, based on \eqref{Eqn:iterSEb}, we address this issue by analyzing the first fixed point $(\rho^*, v^*)$ of the SE of CD-OAMP:
\BE\label{Eqn:iterSE}
    \rho^{*} = {\gamma}_{\mr{SE}}(v^*,\bm{p}), \quad
    v^* = {\phi}_{\mr{SE}}(\rho^{*}).
\EE

Based on this, the problem $\mathcal{P}_{1.1}$ can be numerically solved by determining $v^*=v_{\rm goal}$ such that the objective function $\gamma_{\mr{SE}}(v,\bf{p})-\phi_{\mr{SE}}^{-1}(v)$ is equal to zero. Hence, we can use bisection search to find $v^*$. It is difficult to find the minimum  over the continuous interval ${v \in [v_{\rm goal}, 1)}$. To address this, we evaluate it at uniformly sampled points $\mathcal{V}_{\rm goal} = \{v_i\}$ in the logarithmic domain over the interval $[v_{\rm goal},1)$. Typically, we use $100$ sampling points, i.e., $|\mathcal{V}_{\rm goal}| = 100$.

In summary, we solve $\mathcal{P}_{1.1}$ numerically by using a bisection search to find $v_{\rm goal}$ such that the objective function of the following $\mathcal{P}_{1.3}$ is equal to zero:
\BS \label{Eqn:P2_b} \begin{align} 
\!\! {\mathcal P}_{1.3}: \;\; \mathop{\rm{max}}_{\bm{p}} \; \mathop{\rm{min}}_{v \in \mathcal{V}_{\rm goal}} &\;  \gamma_{\mr{SE}}(v,\bf{p})-\phi_{\mr{SE}}^{-1}(v),\\
{\rm s.t.}\;&     \;\;  \textstyle\sum_{i=1}^N p_i =P_{\rm sum},\\
& \;\; p_i\geq0,\; i=1,\cdots\!, N.
\end{align}\ES
Since $\gamma_{\mr{SE}}(v,\bf{p})$ in \eqref{Eqn:P2_b} is concave w.r.t $\bf{p}$ and the constraints are convex, ${\mathcal P}_{1.3}$ is a convex problem, allowing us solving it by using standard convex optimization solvers.

\subsection{Power Allocation to Maximize Constrained Capacity}\label{Sec:PA_Cap}
We now study the power allocation to maximize the constrained capacity of the system in \eqref{Eqn:pa2}, which is an issue not considered in \cite{EST-EQ}. The constrained capacity can be evaluated by the following lemma.
\begin{lemma}[Replica Constrained Capacity]\label{Pro:Cap_PA}
     The replica constrained capacity per transmit symbol of the multiple-input multiple-output (MIMO) linear system in \eqref{Eqn:pa2} is given by
\BE\label{Eqn:C_MIMO_PA}
C_{\rm MIMO}(\bm{p}) =  \int_0^{v^*{\rm snr}} \!\!\!\!\!\!\!\!\!\! \mathcal{R}_{\bf{R}}(-z)dz + C_{\rm SISO}(\rho^*) - \rho^*v^*,
\EE
where $\bf{R}=\bm{\Sigma}_P^{\rm H}\tilde{\bm{\Sigma}}_{A}^{\rm H}\tilde{\bm{\Sigma}}_{A}\bm{\Sigma}_P$, $\rho^*={\rm mmse}^{-1}(v^*)$, and $v^*$ is the replica MMSE given in \eqref{Eqn:replicaMMSE_PA}. 
\end{lemma}     

The power allocation to maximize the constrained capacity is formulated as the following optimization problem:
\BS\label{Eqn:optP_rate}
\begin{align}
      {\mathcal P}_{2}:  \;\;    &\mathop{\rm{max}}_{\bm{p}}  \;\;  C_{\rm MIMO}(\bm{p}),\\
     & {\rm s.t.}\quad  \textstyle\sum_{i=1}^N p_i = P_{\rm sum}, \\ 
     & \;\;\quad\quad p_i\geq 0,\; i=1,\cdots\!, N,
\end{align}\ES 
where $C_{\rm MIMO}(\bm{p})$ is given in \eqref{Eqn:C_MIMO_PA}. 

For Gaussian signaling $\bf{s}$, $C_{\rm MIMO}$ simplifies to the Gaussian MIMO capacity $C_{\rm Gau-MIMO}(\bm{p}) = \frac{1}{N} \sum_{i=1}^N \log\left(1 + p_i \varrho_i\right)$, where $\varrho_i=\sigma_i^2/\sigma^2$. In this case, the optimal power allocation corresponds to the \emph{water-filling} solution in \cite[Sec. 10.4]{Cover1990}. Conversely, for non-Gaussian signaling $\bf{s}$, the power allocation becomes more complex. Following the capacity-area theorem in \cite[Theorem 1]{LeiOptOAMP}, the linear capacity $C_{\rm MIMO}$ in \eqref{Eqn:C_MIMO_PA} can be reformulated to
\BS\label{Eqn:TF}\BE
    C_{\rm MIMO} (\bf{p}) = \int_{0}^1 \min\big\{{\eta}_{\rm SE}(v,{\bm p}), {\rm mmse}^{-1}(v)\big\} d v,
\EE
where ${\eta}_{\rm SE}(\cdot)$ is the variational transform function given by
\begin{align}
    {\eta}_{\rm SE}(v,\bf{p})\equiv v^{-1}-[\hat{\gamma}_{\rm SE}^{-1}(v,{\bm p})]^{-1}, \label{Eqn:eta_p}
\end{align}\ES
where $\hat{\gamma}^{-1}_{\rm SE}(v,\bm{p})$ denotes the inverse function of $v=\hat{\gamma}_{\rm SE}(\tilde{v},\bm{p})$ w.r.t. $\tilde{v}$.
It is important to note that the ${\phi}_{\mr{SE}}(\cdot)$ in \eqref{Eqn:phi_se} involves an orthogonalization operation in the SE of CD-OAMP, rendering it no longer locally MMSE optimal. This complicates the analysis of constrained capacity and achievable rates using the I-MMSE lemma \cite{Guo2005}. This difficulty can be overcame by utilizing the variational transform functions of CD-OAMP, as outlined in \cite[Equation (37)]{LeiOptOAMP}, which preserves the same fixed point as the SE described in \eqref{Eqn:iterSEb}. Moreover, since $[\hat{\gamma}_{\rm SE}^{-1}(v,{\bm p})]^{-1}$ in \eqref{Eqn:TF} is convex w.r.t $\bf{p}$, ${\mathcal P}_{2}$ is a convex optimization problem. As a result, we can solve it by using standard convex optimization solvers.

\section{Numerical Results}
We consider that the carrier frequency is $4$~GHz with $\Delta f = 15$ kHz, the velocity of the device is $v=300$ km/h with a maximum Doppler frequency shift $\nu_{\text{max}}=1111$~Hz, the maximal number of multipaths is $5$, and the channel Doppler shift is generated by using Jakes information \cite{OTFS_GMP}. The root raised-cosine filter rolloff factor employed in the transceiver is set at $0.4$. Here, time-varying SISO and MIMO random modulation systems are considered with transceiver antenna $(J, K)=(1,1)$, $(8, 4)$, and $(4, 4)$, where correlation parameter $\rho=\{0, 0.6\}$ and $M=N=\{2048, 256\}$, respectively. 

\begin{figure}[t!]\vspace{-0.5cm}
    \centering
    \includegraphics[width = 0.65\columnwidth]{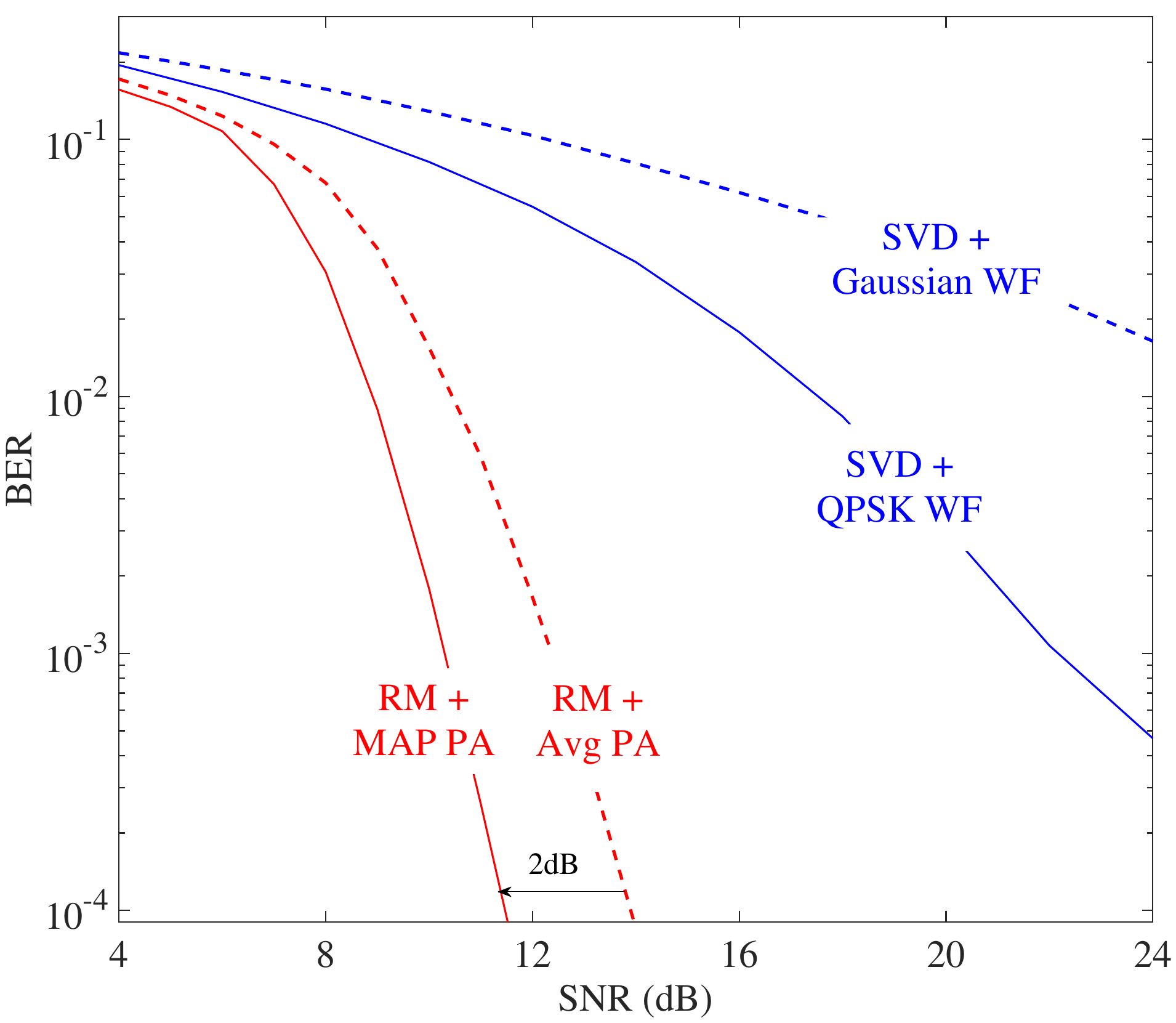}\vspace{-0.2cm}
    \caption{BER of random modulation (RM) with CD-MAMP detector in SISO systems with QSPK signaling and $M=N=2048$, where MAP PA denotes optimal-MAP power allocation  \eqref{Eqn:P2_b}.}
    \label{fig:RM_PA}
\end{figure}
\begin{figure}[t!]\vspace{-0.5cm}
    \centering
    \includegraphics[width = 0.65\columnwidth]{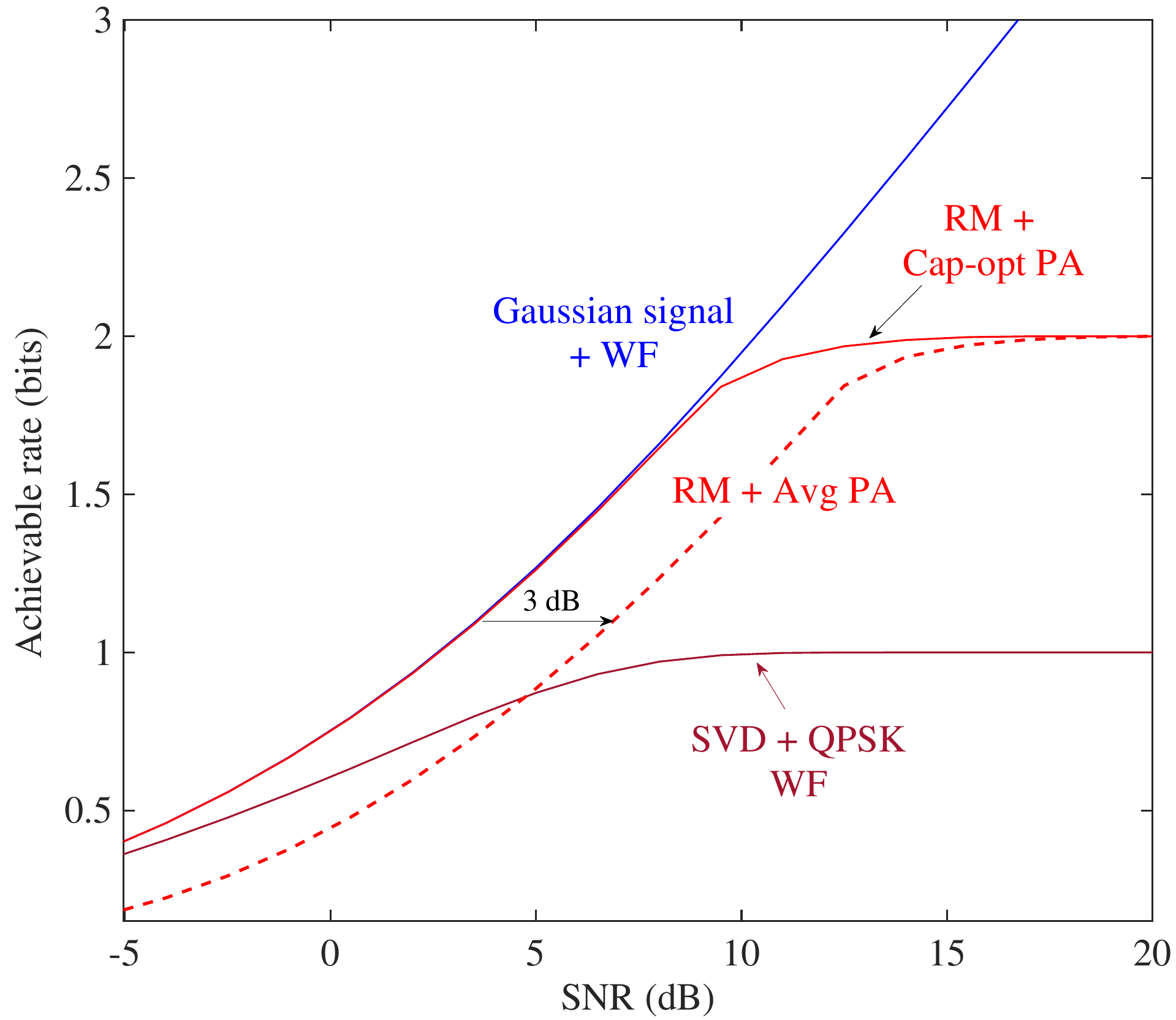} \vspace{-0.2cm}
    \caption{Maximum achievable rate of CD-MAMP with 
    random modulation (RM) in MIMO linear systems, where QPSK signaling, and different power allocation (PA) (i.e. optimization in \eqref{Eqn:optP_rate} and water-filling (WF) are employed with $M=N=256$ and $(J=8, N=4)$.}
    \label{fig:MIMO_qpsk}
\end{figure}

\begin{figure}[t!]
    \centering
    \includegraphics[width = 0.65\columnwidth]{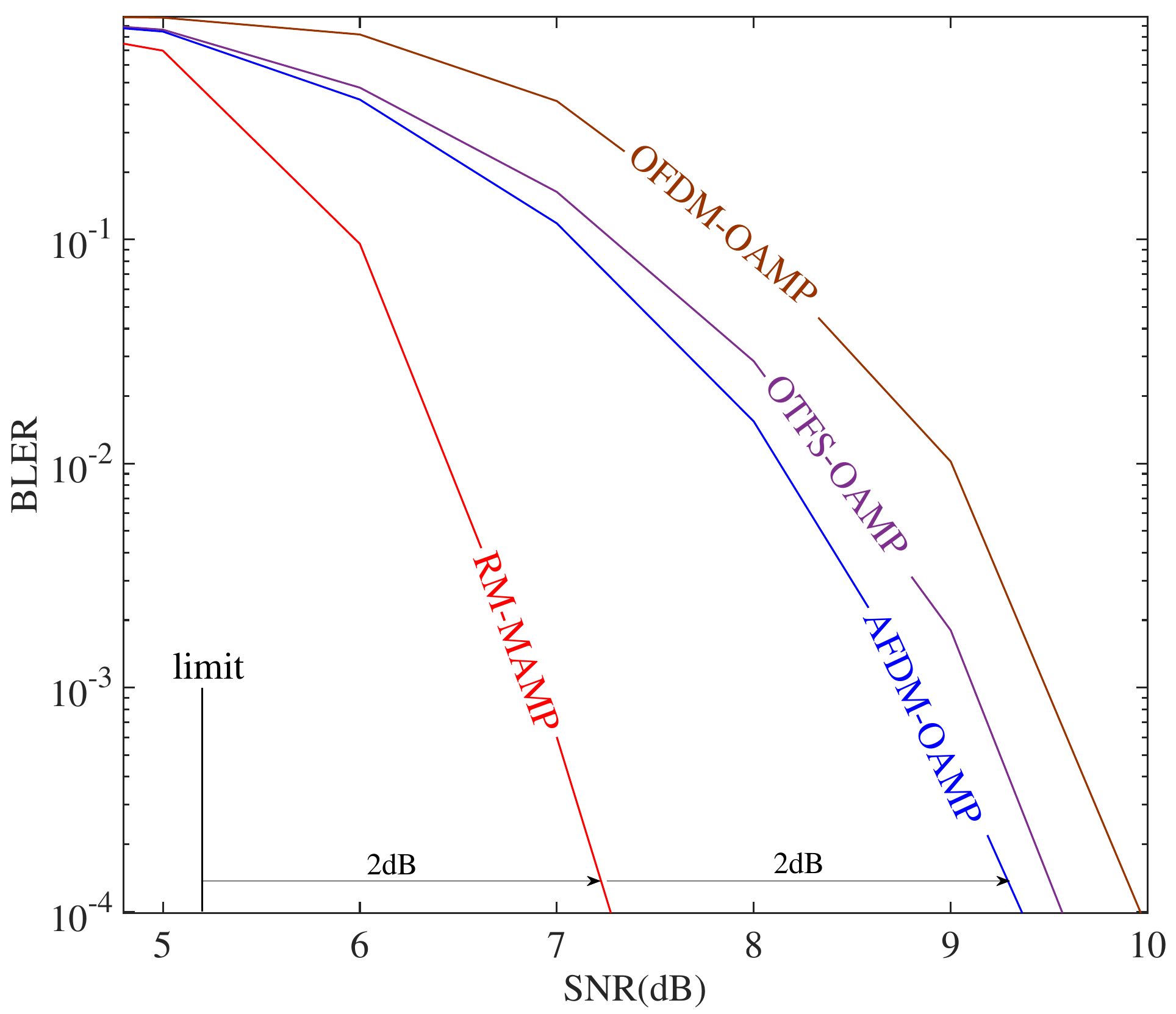}\vspace{-0.2cm}
    \caption{BLER performances of CD-MAMP and CD-OAMP (abbreviated as MAMP and OAMP) with 5G-NR LDPC codes in MIMO systems with coding rate is $0.625$, $(J, K)=(4,4)$ and $M=N=2048$, where random modulation (RM), OFDM, OTFS, and AFDM are employed with QPSK signaling.}
    \label{fig:MIMO_SIC}
\end{figure}

Fig.~\ref{fig:RM_PA} and Fig.~\ref{fig:MIMO_qpsk} show the BER and achievable rate comparisons of random modulation with optimal-MAP power allocation (PA) in \eqref{Eqn:P2_b}, average PA, and channel parallelization via singular value decomposition (SVD)\cite{Telatar1999}  with Gaussian and QPSK water filling, where QSPK signaling is employed. In random modulation, the optimal power allocation, aimed at achieving MAP performance and maximum achievable rate, can achieve a $2\sim 3$ dB gain over average power allocation, respectively. In contrast, water-filling for Gaussian and QPSK signaling using channel parallelization via SVD decomposition results in a significant performance loss of more than $12$~dB and almost half rate loss in the high SNR region (i.e., SNR $> 10$ dB).  Fig.~\ref{fig:MIMO_SIC} shows the BLER comparisons of the different modulation with 5G-NR LDPC codes in MIMO linear systems, where the coding rate is $0.625$ and coding length is $2048$. Note that the gap between the BER curve of random modulation with CD-MAMP at $10^{-4}$ and the corresponding performance limit is $2$ dB, which achieves up to a $2$ dB gain with lower complexity compared to OTFS, AFDM, and OFDM with CD-OAMP.

\section{Conclusion}
This paper presents the random modulation for arbitrary norm-bounded and spectrally convergent channels. By employing a random transform, the proposed CD-MAMP detector can achieve the asymptotic replica MAP-BER optimality with low complexity, which effectively utilizes the sparsity of time-domain channels. Furthermore, the optimal power allocation strategies are developed to minimize the MAP BER and maximize the constrained channel capacity. Numerical results verify the advantages of proposed random modulation over existing OFDM/OTFS/AFDM schemes.
\bibliographystyle{IEEEtran}
\bibliography{reference}
\end{document}